# Thermal Evolution of Uranus with a Frozen Interior

**Authors:** Lars Stixrude[1*], Stefano Baroni[2,3], Federico Grasselli[2]

**Affiliations:**

[1]Department of Earth, Planetary, and Space Sciences, University of California, Los Angeles, CA 90095.

[2]SISSA-Scuola Internazionale Superiore di Studi Avanzati, Trieste, Italy.

[3]CNR-IOM DEMOCRITOS@SISSA, Trieste, Italy.

Correspondence to: lstixrude@epss.ucla.edu.

**Abstract:** The intrinsic luminosity of Uranus is a factor of 10 less than that of Neptune, an observation that standard giant planetary evolution models, which assume negligible viscosity, fail to capture. Here we show that more than half of the interior of Uranus is likely to be in a solid state, and that thermal evolution models that account for this high viscosity region satisfy the observed faintness of Uranus by storing accretional heat deep in the interior. A frozen interior also explains the quality factor of Uranus required by the evolution of the orbits of its satellites.

**One Sentence Summary:** We propose that the interior of Uranus is largely frozen and that the presence of a frozen core explains the anomalously low heat flow of the planet, and the tidal dissipation required by the orbits of its moons.

**Main Text:** Despite their similarity in mass and radius, Uranus and Neptune have remarkably different thermal histories (*1*). Whereas homogeneously convecting, inviscid thermal evolution models capture the present day intrinsic luminosity of Neptune, that predicted for Uranus is much too large, and the time scale required for cooling to the present state is much longer than the age of the solar system ($\tau$=8-10 Gyr). Another puzzling observation is that tidal dissipation of Uranus is much larger than expected for an entirely fluid planet.

It has been shown that the thermal evolution of Uranus can be reconciled with observations by trapping heat in its interior (*2*). Proposed mechanisms for trapping this heat focus on reducing the efficiency of thermal convection in the deep interior, for example by the presence of compositional gradients that compete with thermal buoyancy, producing stagnant, double diffusive, or turbulently diffusive layers (*3, 4*). These scenarios invoke physical processes that are difficult to constrain.

Here we show that the thermal evolution and tidal dissipation of Uranus can be explained by considering the effects of freezing. Experimental measurements show that the melting temperature of $H_2O$ is much higher at high pressure than previously thought (*5, 6*). According to experimental determinations of the phase diagram and current models of the thermal structure of Uranus, more than half of the interior of Uranus may be composed of a solid superionic phase, in which a molten sub-lattice of H atoms exists within a crystalline sub-lattice of oxygen atoms. It has been argued that the viscosity of this superionic phase is very similar to that of liquid water, in which case the freezing transition has essentially no influence on thermal evolution (*7*). However, we find that the arguments used to estimate the viscosity of the superionic phase are inappropriate to the solid state.

We find that the viscosity of the superionic phase is dominated by the crystalline oxygen sublattice (8). Previous first principles simulations used the shear stress auto-correlation function to determine the viscosity, finding a very small value, comparable to liquid water (7). However, this approach is only appropriate in materials that do not support shear stress in strained configurations. We find that superionic ice does support shear stress, and we find values of the elastic constants in good agreement with previous determinations by ab initio molecular dynamics simulations at high pressure and temperature (9).

In the presence of a viscous frozen core, thermal evolution is governed by the equations (1, 10)

$$L_{int} = L_{fluid} + L_{core} = 4\pi R^2 \sigma \left(T_{eff}^4 - T_{eq}^4\right) \tag{1.1}$$

$$L_{fluid} = -\int_{M_c}^{M} C_P \frac{\partial T}{\partial t} dm \tag{1.2}$$

$$L_{core} = -\int_{0}^{M_c} C_P \frac{\partial T}{\partial t} dm - \int_{S} C_P \Delta T \frac{\partial c}{\partial t} \rho_c \, dS \tag{1.3}$$

Where $L_{int}$ is the total luminosity from the interior, $L_{fluid}$ and $L_{core}$ are the contributions from the fluid envelope and the frozen core, respectively, $R$ is the radius of the planet, $\sigma$ is the Stefan-Boltzmann constant, $T_{eq}$ is the luminosity due to thermalized and re-radiated solar flux, $T_{eff}$ is the temperature the planet would have in the absence of solar luminosity, $M$ is the mass of the planet, $M_c$ is the mass of the frozen core, $C_P$ is the isobaric specific heat, $c$ is the radius of the frozen core, $S$ is the surface of the core, $\rho_c$ is the density at the core radius, $\Delta T$ is the temperature contrast across the thermal boundary layer at the top of the core, $\partial T/\partial t$ is the cooling rate, and $\partial c/\partial t$ is the growth rate of the core.

The system of equations differs from the standard thermal evolution model for giant planets in the appearance of a thermal boundary layer at the top of the core [second term on the right hand side of Eq. (1.3)]. As $\partial c/\partial t$ and $\Delta T$ are positive quantities, this equation shows that the presence of a thermal boundary layer reduces $L_{core}$, as compared with a uniformly fluid planet, thereby storing heat in the deep interior and allowing the fluid envelope to cool more quickly. In the case of a purely fluid planet, $\Delta T \to 0$, and the interior heat flux reduces to that of the homogeneous case, i.e. $L_{int} = -\int_{0}^{M} C_P \partial T/\partial t \, dm$ (1).

The thermal evolution equations are closed by adopting a relationship between the core luminosity and the temperature contrast (11)

$$L_{core} \propto k \left(\frac{\rho \alpha g}{\eta \kappa}\right)^{1/3} \Delta T^{4/3} \tag{1.4}$$

which shows that the core luminosity increases with the thermal conductivity $k$, and the temperature contrast. The other parameters are thermal expansivity $\alpha$, thermal diffusivity $\kappa = \rho C_P k$, gravitational acceleration $g$, and viscosity $\eta$. We determine the relevant physical parameters from first-principles molecular dynamics simulations (*12*). For the thermal conductivity, we use the Green-Kubo theory of linear response, leveraging a recently developed "gauge invariant" approach to heat transport (*13*) in multi-component systems (*14*), and the resulting cepstral analysis of the Onsager coefficients (*14, 15*). For the viscosity, we adopt the homologous temperature relationship that has been widely used in studies of the thermal evolution of icy moons and is supported by experimental data (*16, 17*). The inner envelope of Uranus is unlikely to be composed of pure $H_2O$ (although $H_2O$ may be the dominant component), and the presence of impurities may influence the melting temperature and therefore the viscosity. There has been one study performed on impure $H_2O$ systems at the relevant conditions, which found that a 1:1 mix of water and ammonia had nearly the same melting point as pure water (*18*).

We solve equations (1.1)-(1.4) for Uranus with a fourth-order Runge-Kutta scheme (*8*). We adopt planetary structure models that agree with observational data (*19*). In order to focus on the effect of finite viscosity, we simplify the description of thermodynamic properties by assuming homogeneous values of the heat capacity $C_P$. Away from the thermal boundary layer, we assume that the temperature gradient is adiabatic, and that the adiabatic gradient $\nabla = \left(d \ln T / d \ln P\right)_S$ is homogeneous. We ignore the relatively minor contributions to luminosity from gravitational contraction (*2*). We do not explicitly account for compositional layering in Uranus: we assume that over the range of depth that the core forms in our models (out to 2/3 of the planet's radius) the composition is dominated by $H_2O$, in accord with detailed interior models (*20*). We neglect the effects of an innermost rocky layer, and the latent heat of freezing on thermal evolution.

Our results show that the presence of a frozen core explains the observed heat flow of Uranus (Fig. 1). The cooling time scale decreases by nearly a factor of two as compared with the homogeneous case. The frozen-core evolution coincides with the homogeneous evolution over the first 1 Gyr while the core is absent or still too small to significantly affect the luminosity. As the core grows, the effective temperature drops more rapidly as the core retains an increasing proportion of the internal heat.

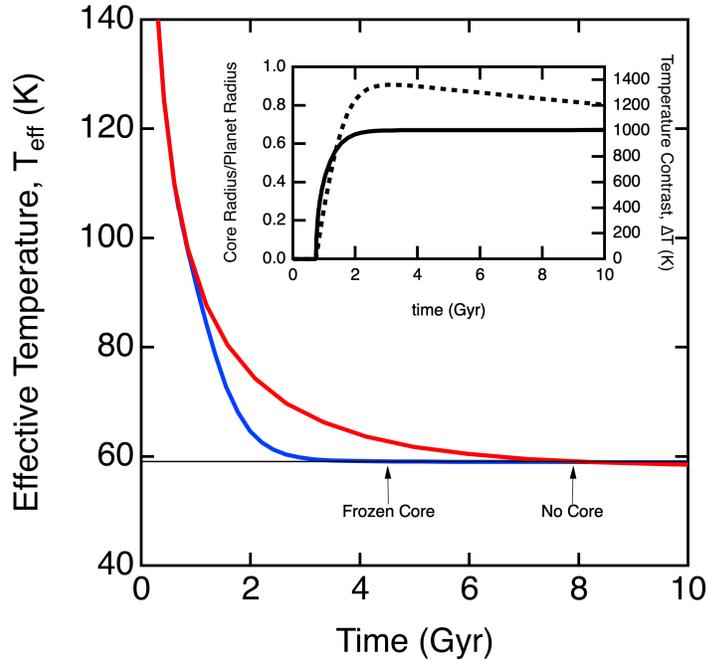

**Fig. 1.** Our thermal evolution model of Uranus containing a frozen core (blue line) as compared with a model that is homogeneous and completely fluid (red line). The arrows indicate the time at which the effective temperature reaches the observed, present-day value (59.1 K). Inset: The internal structure of the frozen core model showing the evolution of the core radius $c$ (solid line, left-hand axis) and the temperature contrast at the top of the core $\Delta T$ (dashed line, right-hand axis). We assume $C_P$=5000 J kg$^{-1}$ K$^{-1}$, and $\nabla$=0.2585.

The planet starts off in an entirely fluid state as the temperature is everywhere greater than the melting point of superionic H$_2$O. The core first nucleates at ~0.8 Gyr when the adiabat intersects the freezing curve. The core grows rapidly initially because cooling is more rapid at earlier times and because of the curvature of the freezing transition: the Clapeyron slope $(dT/dP)_{eq}$ is smaller at high temperature (Figs. 1,2). The temperature contrast at the top of the core grows with time initially. After ~2 Gyr, the core radius and the temperature contrast show little further change.

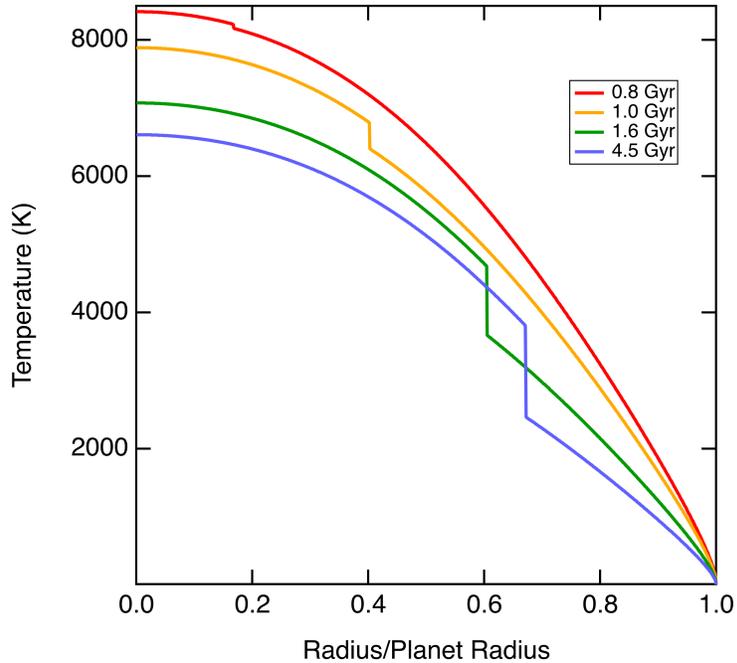

**Fig. 2.** Temperature profiles in the frozen core model shown in Fig. 1 at different times as indicated.

We find that a wide range of values of the heat capacity and the thermal gradient yield cooling times in agreement with observation (*8*). Because the composition of the interior cannot be uniquely constrained by current observations, we explore the range $C_P$=3000-6000 J kg$^{-1}$ K$^{-1}$, which encompasses the Dulong-Petit values of pure $H_2O$, a $H_2O$-$NH_3$-$CH_4$ solar mix (*21*), and the possible effects of admixtures of hydrogen, helium, and heavy elements, and the range $\nabla$=0.24-0.29, which encompass detailed multi-layer compositional models of Uranus (*20, 22*). A recent study argued that the cooling time for Uranus may be shorter than 4.5 Gyr and that the planet has been in radiative equilibrium for some time, because present observations show $T_{eff}$ and $T_{eq}$ are indistinguishable within uncertainty (*3*). Our frozen core model can also accommodate cooling times as short as 1 Gyr over the range of $C_P$ and $\nabla$ that we have explored (*8*).

The presence of a frozen core in Uranus can also explain the large tidal dissipation that is required by the orbits of its satellites (Fig. 3). Analyses of orbital evolution requires that the tidal quality factor $Q$ (inverse of dissipation) is much smaller than expected from the effects of turbulent viscosity in an entirely fluid planet (*23-25*). A frozen core provides a natural mechanism for producing the required dissipation. We model the frozen core as a standard linear solid, and find agreement with the astronomical constraints for plausible parameter values (Fig. 3) (*8*). In the initial stages of evolution, the value of $Q$ is very large because the planet is entirely fluid. As the core begins to grow, the value of $Q$ diminishes. The quality recovers as the core becomes very large and an increasing proportion of it becomes so cold that it no longer contributes significantly to dissipation.

Our thermal evolution model is in excellent agreement with the observed characteristics of the orbits of Uranian moons. We have computed the evolution of the radii of the orbits of the Uranian satellites using the time-varying value of the tidal quality factor from our thermal evolution model (*8*). The evolution of the orbital radii is initially very slow because $Q$ is very high. As $Q$ decreases, the orbits evolve more rapidly. We find that Miranda:Umbriel pass through a 3:1 resonance, which explains the anomalously high inclination of Miranda (*25*), while Ariel:Umbriel avoid a 2:1 resonance, in which they would have remained had they ever encountered it, contrary to observations (*25*), and Miranda:Ariel avoid a 5:3 resonance which would have caused a large, unobserved increase in the eccentricity of Miranda's orbit.

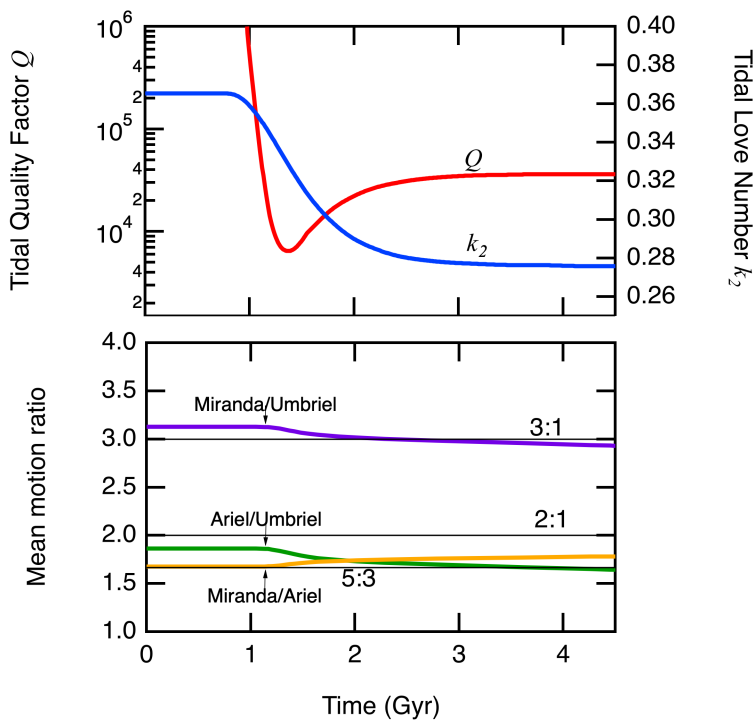

**Fig. 3**. (top) Tidal quality factor (red line, left-hand axis) and $k_2$ (blue line, right-hand axis) of Uranus as a function of time for the thermal evolution model shown in Fig. 1 and a standard linear solid model of viscoelasticity. (bottom) the ratio of the mean motion of the indicated satellite pairs (colored lines) as compared with the 2:1, 3:1, and 5:3 resonances (*25*).

Our thermal evolution model makes predictions that can be tested by future observational missions. Because of the presence of a frozen core, the tidal Love number $k_2$ is much less than that of a fluid body (Fig. 3). We find $k_2$=0.275 for our model at the present day, whereas we calculate $k_2$=0.363 for the same density model without rigidity. The difference between fluid and frozen results is much larger than the uncertainty on $k_2$ of other planets for which this quantity has been measured. The tidal love number has not yet been measured for Uranus, although the possibility of a future measurement has been discussed (*26*).

Frozen cores may exist in other giant planets. In the case of Neptune, purely fluid thermal evolution models readily account for the observed cooling time (*1*). However, this does not preclude the presence of a frozen core and it may be possible to find frozen core models that explain the thermal evolution of Neptune. Such a model may help to explain the tidal $Q$ of Neptune which, like Uranus, must be much less than the purely fluid value in order to explain the characteristics of its satellite orbits. Frozen icy cores may also exist in ice-rich exoplanets and this possibility should be considered in evaluating the evolution of orbits about host stars (*27*).

The presence of a frozen core is important for understanding the generation of magnetic fields in icy bodies. The frozen core is far too viscous to produce a dynamo. The dynamo must instead be produced in the upper fluid third of the radius of the planet. Previous simulations have investigated thin shell dynamos and find that the thickness of the shell has an important effect on the geometry of the field (*28, 29*).

**REFERENCES AND NOTES**

**Acknowledgments:** FG thanks R. Bertossa and L. Ercole for technical support on the data analysis of time series. **Funding:** This project is supported by the National Science Foundation under grant EAR-1853388 and by the EU through the MaX Centre or Excellence for Supercomputing Applications (Projects No. 676598 and 824143). **Author contributions:** LS FG and SB designed and performed the research and wrote the paper. **Competing interests:** The authors declare no conflict of interest. **Data and materials availability:** All data are available in the manuscript and supplementary materials.

**Supplementary Materials:**

Materials and Methods

Figures S1-S5

Table S1-S2

References (*30-61*).

**REFERENCES AND NOTES**

1. J. J. Fortney, N. Nettelmann, The Interior Structure, Composition, and Evolution of Giant Planets. *Space Sci Rev* **152**, 423-447 (2010).
2. W. B. Hubbard, M. Podolak, D. J. Stevenson, in *Neptune and Triton,* D. Cruishank, Ed. (University of Arizona Press, Tucson, 1995), pp. 109-138.
3. N. Nettelmann *et al.*, Uranus evolution models with simple thermal boundary layers. *Icarus* **275**, 107-116 (2016).
4. M. Podolak, R. Helled, G. Schubert, Effect of non-adiabatic thermal profiles on the inferred compositions of Uranus and Neptune. *Monthly Notices of the Royal Astronomical Society* **487**, 2653-2664 (2019).
5. M. Millot *et al.*, Experimental evidence for superionic water ice using shock compression. *Nat. Phys.* **14**, 297-+ (2018).
6. M. Millot *et al.*, Nanosecond X-ray diffraction of shock-compressed superionic water ice. *Nature* **569**, 251-255 (2019).
7. R. Redmer, T. R. Mattsson, N. Nettelmann, M. French, The phase diagram of water and the magnetic fields of Uranus and Neptune. *Icarus* **211**, 798-803 (2011).
8. Materials and Methods are available as supplementary materials.
9. J. A. Hernandez, R. Caracas, Superionic-Superionic Phase Transitions in Body-Centered Cubic H2O Ice. *Physical Review Letters* **117**, 5 (2016).

10. S. Cottaar, B. Buffett, Convection in the Earth's inner core. *Physics of the Earth and Planetary Interiors* **198**, 67-78 (2012).
11. G. Schubert, D. L. Turcotte, E. R. Oxburgh, Stability of planetary interiors. *Geophysical Journal of the Royal Astronomical Society* **18**, 441-& (1969).
12. F. Grasselli, L. Stixrude, S. Baroni, Heat transport in water at icy-giants conditions from ab initio MD simulations. *arXiv*, 2003.12557 (2020).
13. A. Marcolongo, P. Umari, S. Baroni, Microscopic theory and quantum simulation of atomic heat transport. *Nat. Phys.* **12**, 80-U111 (2016).
14. L. Ercole, A. Marcolongo, S. Baroni, Accurate thermal conductivities from optimally short molecular dynamics simulations. *Sci Rep* **7**, 11 (2017).
15. R. Bertossa, F. Grasselli, L. Ercole, S. Baroni, Theory and Numerical Simulation of Heat Transport in Multicomponent Systems. *Physical Review Letters* **122**, 6 (2019).
16. T. Spohn, G. Schubert, Oceans in the icy Galilean satellites of Jupiter? *Icarus* **161**, 456-467 (2003).
17. W. B. Moore, Thermal equilibrium in Europa's ice shell. *Icarus* **180**, 141-146 (2006).
18. M. Bethkenhagen, D. Cebulla, R. Redmer, S. Hamel, Superionic Phases of the 1:1 Water-Ammonia Mixture. *J Phys Chem A* **119**, 10582-10588 (2015).
19. R. Helled, J. D. Anderson, M. Podolak, G. Schubert, Interior models of uranus and neptune. *Astrophysical Journal* **726**, 7 (2011).
20. N. Nettelmann, R. Helled, J. J. Fortney, R. Redmer, New indication for a dichotomy in the interior structure of Uranus and Neptune from the application of modified shape and rotation data. *Planet Space Sci* **77**, 143-151 (2013).
21. M. Asplund, N. Grevesse, A. J. Sauval, P. Scott, in *Annual Review of Astronomy and Astrophysics, Vol 47,* R. Blandford, J. Kormendy, E. VanDishoeck, Eds. (Annual Reviews, Palo Alto, 2009), vol. 47, pp. 481-522.
22. M. Bethkenhagen *et al.*, Planetary Ices and the Linear Mixing Approximation. *Astrophysical Journal* **848**, 9 (2017).
23. P. Goldreich, P. D. Nicholson, Turbulent viscosity and jupiters tidal Q. *Icarus* **30**, 301-304 (1977).
24. W. B. Hubbard, Tides in giant planets. *Icarus* **23**, 42-50 (1974).
25. W. C. Tittemore, J. Wisdom, Tidal evolution of the Uranian satellites .3. Evolution through the Miranda-Umbriel 3-1, Miranda-Ariel 5-3, and Ariel-Umbriel 2-1 mean-motion commensurabilities. *Icarus* **85**, 394-443 (1990).
26. V. Lainey, Quantification of tidal parameters from Solar System data. *Celest. Mech. Dyn. Astron.* **126**, 145-156 (2016).
27. W. G. Henning, T. Hurford, Tidal heating in multilayered terrestrial exoplanets. *Astrophysical Journal* **789**, 27 (2014).
28. S. Stanley, J. Bloxham, Numerical dynamo models of Uranus' and Neptune's magnetic fields. *Icarus* **184**, 556-572 (2006).
29. K. M. Soderlund, M. H. Heimpel, E. M. King, J. M. Aurnou, Turbulent models of ice giant internal dynamics: Dynamos, heat transfer, and zonal flows. *Icarus* **224**, 97-113 (2013).

# Supplementary Materials for

Thermal Evolution of Uranus with a Frozen Interior

Lars Stixrude, Stefano Baroni, Federico Grasselli

Correspondence to: lstixrude@epss.ucla.edu.

**This PDF file includes:**

Materials and Methods
Figs. S1 to S5
Table S1 to S2

**Materials and Methods**

Thermal Evolution

We solve the system of equations

$$L_{int} = L_{fluid} + L_{core} = 4\pi R^2 \sigma \left( T_{eff}^4 - T_{eq}^4 \right) \quad (0.1)$$

$$L_{fluid} = -\int_{M_c}^{M} C_P \frac{\partial T}{\partial t} dm \quad (0.2)$$

$$L_{core} = -\int_{0}^{M_c} C_P \frac{\partial T}{\partial t} dm - \int_{S} C_P \Delta T \frac{\partial c}{\partial t} \rho_c \, dS \quad (0.3)$$

$$L_{core} = k \left( \frac{\rho \alpha g}{Ra_{cr} \eta \kappa} \right)^{1/3} \left( \frac{h}{c} \right)^{4/3} \Delta T^{4/3} \quad (0.4)$$

Where $L_{int}$ is heat flux from the interior, $L_{fluid}$ and $L_{core}$ are the contributions to this interior heat flux from the fluid envelope and the frozen core, respectively, $R$ is the radius of the planet, $\sigma$ is the Stefan-Boltzmann constant, $T_{eq}$ is the luminosity due to thermalized and re-radiated solar flux, $T_{eff}$ is the temperature the planet would have in the absence of solar luminosity, $M$ is the mass of the planet, $M_c$ is the mass of the frozen core, $C_P$ is the isobaric heat capacity, $c$ is the radius of the frozen core, $S$ is the surface of the core, $\rho_c$ is the density at the core radius, $\Delta T$ is the temperature contrast across the thermal boundary layer at the top of the core, and $\eta$ is the viscosity. The values of the parameters used in the thermal evolution calculations are reported in Table S1.

The viscous scale height $h = -\left( \partial \ln \eta / \partial r \right)^{-1}$ and other quantities in Eq. (0.4) are evaluated at the surface of the core, including the thermal conductivity, $k$, the thermal expansivity, $\alpha$, gravitational acceleration $g$, thermal diffusivity $\kappa$. $Ra_{cr}$ is the critical Rayleigh number for the onset of convection. This expression is appropriate for very deep convective systems, such as the core of Uranus, over which the viscosity varies by many orders of magnitude. We adopt the value $Ra_{cr} = 30$ corresponding to the limit $h/c \to 0(1)$.

The rate of growth of the core is governed by the intersection of the adiabat in the fluid envelope and the phase transition from the fluid to the solid phase (*2*)

$$\frac{\partial c}{\partial t} = -\frac{1}{\rho g (\Gamma - \Gamma_a)} \frac{\partial T_c}{\partial t} \quad (0.5)$$

where $\Gamma$ is the Clausius-Clapeyron slope of the phase boundary, $\Gamma_a = (T_c / P_c) \nabla$ is the slope of the planetary temperature distribution and $T_c$ and $P_c$ are the temperature and pressure at the top of the core. We represent the phase boundary with the Simon equation

$$T_m = T_0 \left(1 + \frac{P - P_0}{a}\right)^b \tag{0.6}$$

and the temperature in the fluid envelope and the solid core by

$$T(P,t) = T_1(t) \left(\frac{P}{P_1}\right)^\nabla \qquad P < P_c \tag{0.7}$$

$$T(P,t) = T_i(t) \left(\frac{P}{P_c}\right)^\nabla \qquad P \geq P_c \tag{0.8}$$

respectively, where $T_i = T_c + \Delta T$, $P_c$ is the pressure at the top of the core, $P_1$=1 bar, and the relationship between $T_1$ and $T_{eff}$ is given by (3)

$$T_1 = B T_{eff}^{1.244} \tag{0.9}$$

With $B$=0.47529 reproducing the present day observed value of $T_1$=76 K for $T_{eff}$=59.1.

We assume that the adiabatic gradient $\nabla$ (away from thermal boundary layers) and the heat capacity are homogeneous. The assumed form of the temperature structure with homogeneous $\nabla$ matches very well multi-layer models in which the adiabatic gradient is given at every point by an assumed composition and equations of state (EOS) of rock, ice, and gas components (Fig. S1).

For the viscosity, we assume the homologous temperature relationship (4, 5)

$$\eta = \eta_0 \exp A \left(\frac{T_m}{T} - 1\right) \tag{0.10}$$

which yields

$$h = \frac{T_c}{\rho g A (\Gamma - \Gamma_a)} \tag{0.11}$$

for the viscous scale height.

For planetary structure, we adopt the so called "empirical" model of Helled et al. (6). This model is specified by a sixth order polynomial of density in radius and matches all relevant observational data. We compute other structural quantities from this polynomial, including pressure, gravitational acceleration and mass as a function of radius. We have checked that our results are not sensitive to the details of planetary structure by using a very different empirical model that includes a dense core equal to 20 % of the planet's mass and fits relevant observational data equally well (7). We find that computed cooling times differ by 10 % between the two models.

Dependence of thermal evolution on the values of $\nabla$ and $C_P$

We have explored a wide range of possible values of $\nabla$ and $C_P$ for the thermal evolution of Uranus (Fig. S2)

The cooling time increases with increasing heat capacity because this increases the thermal inertia of the planet. The cooling time increases with the adiabatic gradient because this increases the temperature of the planet at depth and therefore also increases the thermal inertia. The presence of a frozen core decreases the cooling time as compared with the homogeneous fluid case over most of the range we have explored. The core stores heat and allows the fluid envelope to cool more quickly. For some values of $\nabla$, the presence of a frozen core causes the cooling time to increase slightly. This is because the core, having retained much of its heat, releases it at a time similar to the homogeneous cooling time, thereby increasing the cooling time.

Tidal Dissipation

We characterize tidal dissipation by the quality factor $Q$

$$Q = \frac{|k_2|}{\text{Im}(k_2)} \quad (0.12)$$

where $k_2$ is the tidal love number. The tidal love number depends on the radial structure of density and viscoelasticity and can be determined by solving the following set of six coupled differential equations (8, 9)

$$\dot{y}_1 = \frac{2}{r} y_1 + \frac{l(l+1)}{r} y_2 \quad (0.13)$$

$$\dot{y}_2 = -\frac{1}{r} y_1 + \frac{1}{r} y_2 + \frac{1}{\mu} y_4 \quad (0.14)$$

$$\dot{y}_3 = \frac{4}{r}\left(\frac{3\mu}{r} - \rho g\right) y_1 - \frac{l(l+1)}{r}\left(\frac{6\mu}{r} - \rho g\right) y_2 + \frac{l(l+1)}{r} y_4 - \frac{\rho l(l+1)}{r} y_5 + \rho y_6 \quad (0.15)$$

$$\dot{y}_4 = -\frac{1}{r}\left(\frac{6\mu}{r} - \rho g\right) y_1 + \frac{2(2l^2 + 2l - 1)\mu}{r^2} y_2 - \frac{1}{r} y_3 - \frac{3}{r} y_4 + \frac{\rho}{r} y_5 \quad (0.16)$$

$$\dot{y}_5 = -4\pi G \rho y_1 - \frac{l+1}{r} y_5 + y_6 \tag{0.17}$$

$$\dot{y}_6 = -\frac{4\pi G \rho (l+1)}{r} y_1 + \frac{4\pi G \rho l(l+1)}{r} y_2 + \frac{l-1}{r} y_6 \tag{0.18}$$

where $l$ is the angular order of the deformation ($l=2$ for tidal deformation), $r$ is radius, $\mu$ is (complex) shear modulus, $g$ is gravitational acceleration, and $G$ is the universal gravitational constant. The functions $y_i$, $i=1,6$ are respectively, the amplitude of the radial and tangential deformation, the radial and shear stresses, and the gravitational potential and its radial derivative. The love number is

$$k_l = -y_5 - (r/R)^l \tag{0.19}$$

Fluid layers require special treatment because they do not transmit deformation or stress and couple to other layers only through the gravitational potential (*10, 11*). We have found that we can obtain accurate solutions through fluid layers, and indeed for entirely fluid planets, by the simple expedient of assigning a small but non-vanishing value of the shear modulus to fluid layers ($\mu=10^{-4}\rho g r$). We use the propagator matrix technique to solve Eqs. (0.13)-(0.18) (*9, 12*).

We validate our code for planetary structures consisting of two homogeneous layers, for which the analytical solution (*13*) and numerical solutions (*14*) are available, and for a suite of purely fluid Jupiter models (*15*) (Figs. S3,S4).

Viscoelasticity

We approximate the core as a standard linear solid for which the complex shear modulus (*16*)

$$\mu = \mu_0 - \frac{\delta\mu}{1 + i\omega\tau_\varepsilon}$$

where

$$\delta\mu = \mu_0 \left(1 - \frac{\mu_1}{\mu_0 + \mu_1}\right)$$

and the characteristic time scale

$$\tau_\varepsilon = \frac{\eta}{\mu_0 + \mu_1}$$

where the viscosity is given by Eq. (0.10). The unrelaxed shear modulus $\mu_0$ is taken from ab initio molecular dynamics simulations of (*17*): we used the published values of the full elastic constant tensor to compute the Voigt-Reuss-Hill average shear modulus (*18*) and approximated the pressure and temperature dependence by

$$\mu_0(P,T) = 101 + P/2.41 - T/23.4$$

with pressure in GPa and temperature in K. We adopt a value of $\mu_1 = 60\mu_0$, similar to a recent study of Earth tides (*19*).

We have chosen the standard linear solid as a balance between parametric simplicity and realistic description of the relevant physics. The standard linear solid is a rough approximation to the behavior of real materials, i.e. polycrystals with a variety of grain sizes in part because of the predominance of a single relaxation time, which causes $Q$ to be overly sensitive to frequency. The Maxwell model, to which the standard linear solid reduces in the limit $\mu_1/\mu_0 \to 0$, is even simpler, but shows unrealistically large attenuation at low frequency (*20*). Other more complex models have also been considered in the planetary literature, including the Burgers model (*20*) and the Andrade model (*21*), which agree better with experimental data, but contain additional parameters that are unconstrained at present at the pressure-temperature conditions of the interior of Uranus.

Orbital Dynamics

We determine the evolution of the radii of satellite orbits by numerically integrating (*22*)

$$\frac{1}{a}\frac{da}{dt} = 3\frac{k_2}{Q}n\frac{m}{M}\left(\frac{R}{a}\right)^5 \tag{0.20}$$

backwards in time from present-day observed values. Here $a$ is the radius of the orbit, $m$ is the mass of the moon, $M$ is the mass of the planet, time varying values of $k_2$ and $Q$ are from our thermal evolution model and $n = \sqrt{GM/a^3}$ is the mean motion.

Molecular dynamics simulations

The thermal conductivity, $k$, was extracted from Car-Parrinello *ab initio* molecular dynamics (AIMD) simulations. The results are collected in Ref. (*23*), where the reader can find a thorough description of the methods and parameters employed.

In order to obtain the isobaric specific heat capacity, $C_P$, and thermal expansion coefficient, $\alpha$, as well as the isothermal compressibility, $K_T$, we ran AIMD simulations at fixed number of particles, $N$, pressure, $P$, and temperature, $T$. Simulations parameters were chosen as in Ref. (*23*). All the AIMD simulations were performed using computer codes taken from the QUANTUM ESPRESSO package, v. 6.1 (*24, 25*) and adopting the Perdew–Burke–Ernzerhof (PBE) density functional and Hamann-Schluter-Chiang, as modified by Vanderbilt (*26*) norm-conserving pseudopotentials (downloadable from http://fpmd.ucdavis.edu/potentials/index.htm), which are accurate also at planetary *PT*-conditions (*27*).

The isobaric specific heat and thermal expansion coefficient, as well as the isothermal bulk modulus can be extracted in a single *NPT* simulation from fluctuations as

$$C_P = \frac{1}{k_B T^2 M} \langle \Delta H^2 \rangle_{NPT},$$

$$\alpha = \frac{1}{k_B T^2} \frac{\langle \Delta H \, \Delta V \rangle_{NPT}}{\langle V \rangle_{NPT}},$$

and

$$K_T^{-1} = \frac{\langle \Delta V^2 \rangle_{NPT}}{k_B T \langle V \rangle_{NPT}},$$

where $H$ is the enthalpy and $M$ is the total mass of the system. We employed standard block analysis and error propagation to estimate the uncertainties on $C_P$, $\alpha$, and $K_T$, which are acceptably small for the relatively long *NPT* simulations we ran ($\geq 20$ ps). By employing two *NPT* simulations at different temperatures $T_1$ and $T_2$, but same $P$, a finite-difference method was also devised to obtain estimates of $C_P$ and $\alpha_P$ with lower statistical uncertainty:

$$C_P = \frac{1}{M} \frac{\langle H_2 \rangle - \langle H_1 \rangle}{T_2 - T_1} \bigg|_P,$$

and

$$\alpha = \frac{2}{\langle V_2 \rangle + \langle V_1 \rangle} \frac{\langle V_2 \rangle - \langle V_1 \rangle}{T_2 - T_1} \bigg|_P.$$

We checked the convergence of the results on the parameters of the thermobarostat. The results are reported in Table S2 in which we also report the Grueneisen parameter $\gamma = V \frac{\partial P}{\partial E}\big|_V = \frac{\alpha K_T}{\rho C_P - \alpha_P^2 K_T T}$. Our values are in good agreement with previous AIMD simulations: from the internal energy ($u$) results reported in Table V of (*28*), we find $C_V = \Delta u / \Delta T = 4.78$ J/(gK), at $\rho = 2.5$ g/cm$^3$ and an average $T = 2500$ K; and in Supplementary Table 5 of (*29*), we find: $C_V = 3.98$ J/(gK), at $\rho = 3.753$ g/cm$^3$ and an average $T = 3500$ K.

Response of superionic ice to shear stress

The time-integral of the auto-correlation function of the shear stress

$$\eta_{ACF} = \frac{1}{V k_B T} \int_0^\infty \langle P_{xz}(t) P_{xz}(0) \rangle \, dt$$

does not represent the proportionality coefficient between the stress and the strain rate in viscoelastic materials, like crystals or superionic materials, nor does it dictate the convective motion of large masses of solid, crystalline material, which is instead ruled by defect diffusion.

We ran two *NVE* simulations to extract $\eta$ as in the previous equation, one at $\approx 2500$ K and 175 GPa, where the system is superionic, and one at approximately the same average temperature, but at a much lower pressure ($\approx$ 50GPa), where the oxygen lattice is molten, and the system is in the ionic liquid phase. We obtained $\eta_{ACF} = (1.50 \pm 0.08)$ mPa s for the superionic phase, and $\eta_{ACF} = (1.89 \pm 0.11)$ mPa s for the liquid phase. Even if these values are consistent with those given in (*30*), we do not agree with the conclusion there drawn that "the superionic phase responds almost like a fluid" (verbatim). The similarity of $\eta_{ACF}$ for the superionic and the ionic liquid phases does not indicate their comparable behavior under external shear stress: if it were so, the stress induced by the instantaneous application of an external shear strain to the SI ice system would decay to zero on a timescale related to the time $\tau \approx 100$ fs over which the auto-correlation function of the shear stress is extinguished. We verified that this is not the case by deforming the simulation cell so to include non-diagonal components in the cell tensor (in particular, we introduced a non-vanishing $xz$ and $zx$ components), and monitoring the behavior of the induced shear stress. On the timescale of $\tau$ we observe a decay of the latter to a new equilibrium value, see Fig. S5. No approach to zero of $P_{xz}$ was observed for the superionic phase, not even after several ps of simulation. On the contrary, in the case of ionic liquid water, $P_{xz}$ vanishes at equilibrium. Therefore, in contrast to (*30*), we conclude that the SI phase does not respond to an external shear stress like a fluid.

**Supplementary Figures**

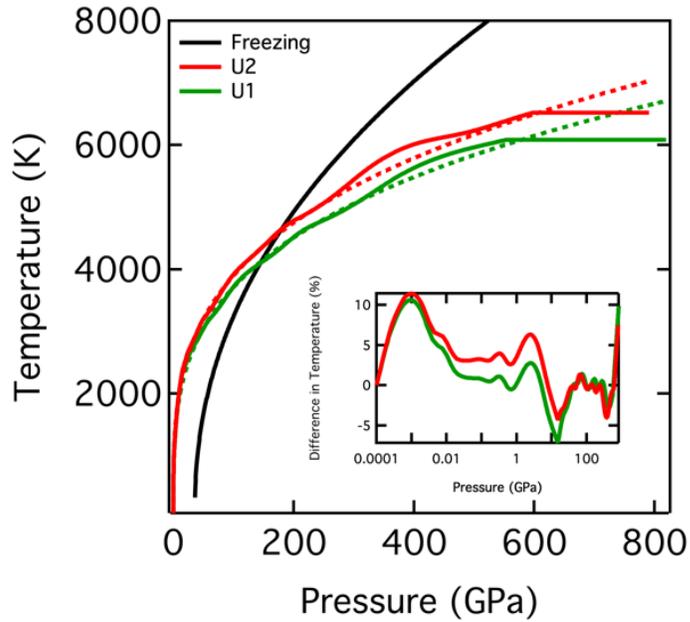

**Fig. S1**. Temperature in multi-layered EOS-based models U1 and U2 (*31*) (solid lines) compared with homogeneous ∇ (dashed lines, ∇=0.28149 for U1 and ∇=0.28505 for U2). The inset shows the difference between the multi-layered and homogeneous ∇ models. Another recent multi-layered model (*32*) (not shown) can be approximated by ∇=0.25507.

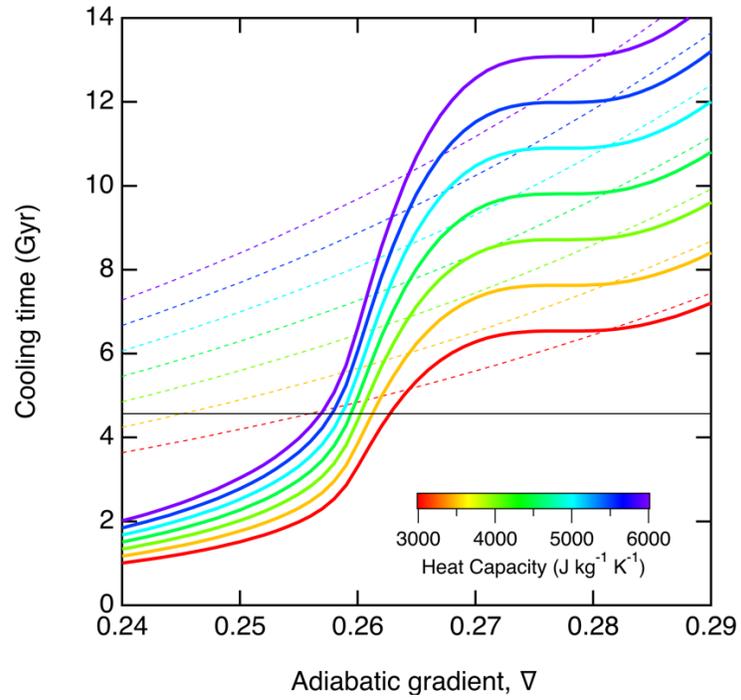

**Fig. S2**. Cooling time of Uranus as a function of the adiabatic gradient for various values of the heat capacity (color bar) for models that contain a frozen core (solid lines) and those that do not (dashed lines).

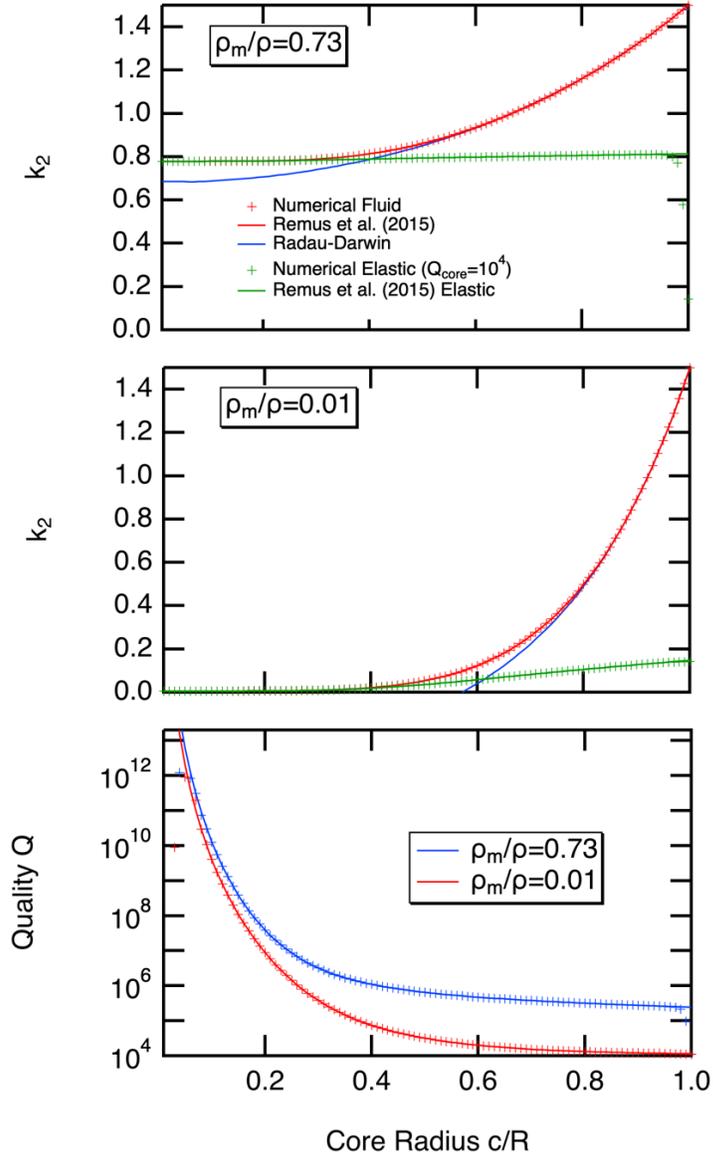

**Fig. S3**. Tests comparing the numerical results of our propagator matrix code (plus symbols) to the analytical solution of Remus et al. (*13*), for planets consisting of two homogeneous layers of density $\rho_m$ and $\rho_c$ and mean density $\rho$ for two different values of $\rho_m/\rho$. Results are shown for purely fluid planets (red in the upper two panels) and planets with a viscoelastic core defined by $Q_{core}=10^4$, a Maxwell rheology and magnitude of the shear modulus $\rho g_{surf} R$ where $g_{surf}$ is the gravitational acceleration at the surface of the planet (green, and lower figure). Note that numerical results capture the correct behavior of the elastic $k_2$ in the limit $c/R \to 1$ (independent of $\rho_m/\rho$) whereas the analytical solution does not obey the correct limit. The numerical solution fails for extremely large values of $Q$ (bottom panel) because of the assumed non-vanishing value of the shear modulus in the fluid layer. In the top two panels the predictions of the Radau-Darwin equation are also shown for comparison. For these tests the mass of the planet $M=6.55 M_{Earth}$, and $R=2.68 R_{Earth}$.

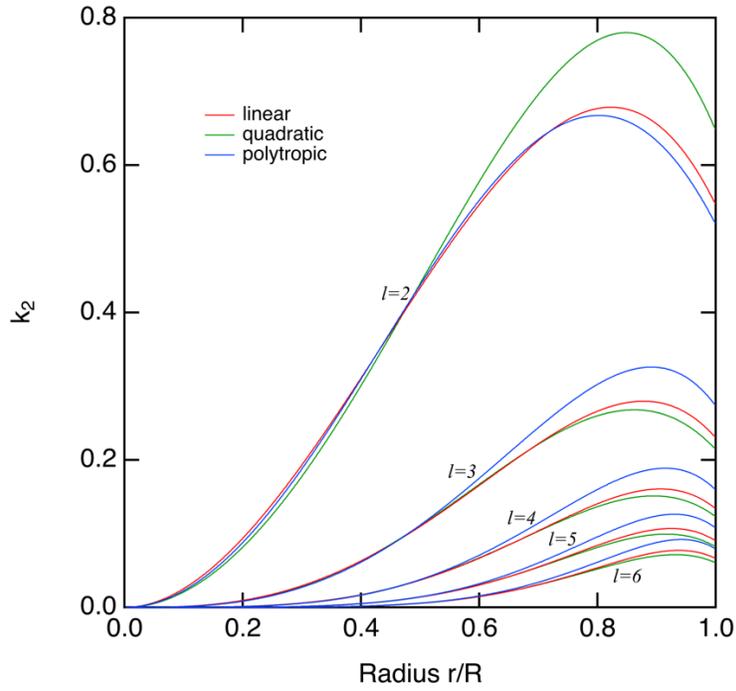

**Fig. S4**. Tidal love number $k_2$ as a function of radius for linear, quadratic, and polytropic models of Jupiter in the fluid limit computed using our propagator matrix code, producing results indistinguishable from those shown in Gavrilov et al. (*15*).

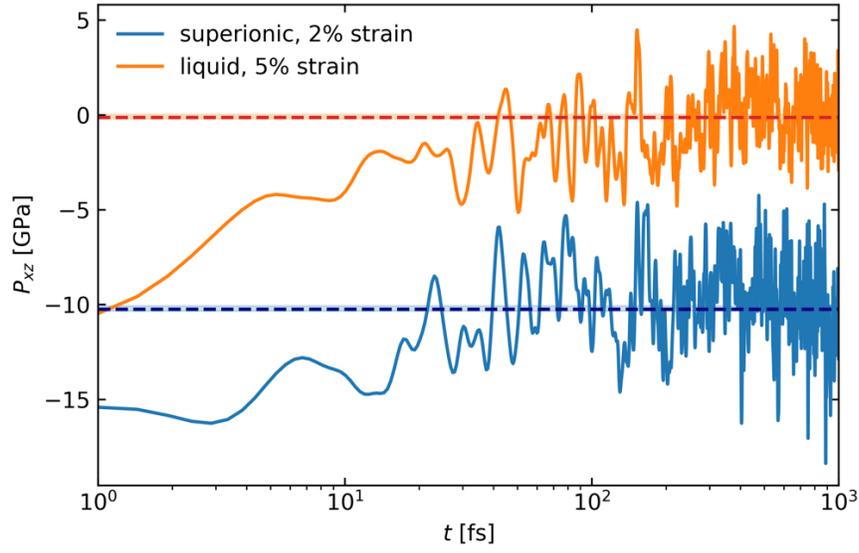

**Fig. S5.** Decay of shear stress $P_{xz}$ in a distorted cell. Blue: superionic ice at $T \approx 2500$ K and $P_{xx} \approx 175$ GPa, with 2% strain on the $xz$ element of the cell tensor. Orange: liquid water at approx. the same temperature but with a diagonal stress $P_{xx} \approx 50$ GPa, and with 5% strain on the $xz$ element of the cell tensor. The dashed horizontal lines are the asymptotic averages computed on longer segments of the simulations from t > 500 fs, i.e. $(-10.2 \pm 0.2)$ GPa and $(-0.09 \pm 0.12)$ GPa, respectively. The errors have been obtained from standard block analysis. The extracted elastic modulus is $c_{44} = 255 \pm 5$ GPa, in excellent agreement with the results of (*17*).

# Supplementary Tables

**Table S1.** Values of parameters used in the thermal evolution model.

| Parameter | | Value |
|---|---|---|
| $R$ | Planet Radius | 25388 km |
| $C_P$ | Specific heat | 3000-6000 J kg$^{-1}$ K$^{-1}$ |
| $\alpha$ | Thermal expansivity | 1.46x10$^{-5}$ K$^{-1}$ |
| $k$ | Thermal conductivity | 11.2 W m$^{-1}$ K$^{-1}$ |
| $T_0$ | Phase Transition reference T | 1000 K |
| $P_0$ | Phase Transition reference P | 40 GPa |
| $a$ | Simon Pressure | 4.6238 GPa |
| $b$ | Simon exponent | 0.44646 |
| $\nabla$ | Adiabatic gradient | 0.24-0.29 |
| $T_{eq}$ | Radiative equilibrium temperature | 58.1 K |
| $\eta_0$ | Reference viscosity | 7x10$^{14}$ Pa s |
| $A$ | Viscosity exponent | 26 |
| $\omega$ | Rotational frequency | 1.012x10$^{-4}$ s$^{-1}$ |
| $\mu_1/\mu_0$ | Standard Linear Solid parameter | 60 |

**Table S2**. Thermodynamic properties of phases: Liquid H$_2$O, which under these conditions is partially dissociated (ionic liquid), and superionic H$_2$O via fluctuations (lines 1-3) and compared with the 2-point method for the superionic phase (line 4).

| phase | $T[K]$ | $P[GPa]$ | $\rho\left[\frac{g}{cm^3}\right]$ | $C_P\left[\frac{J}{gK}\right]$ | $\alpha_P\left[\frac{10^{-6}}{K}\right]$ | $K_T[GPa]$ | $\gamma$ |
|---|---|---|---|---|---|---|---|
| *ionic liquid* | 2000 | 20 | 1.75 ± 0.02 | 4.22 ± 0.13 | 44 ± 6 | 80 ± 6 | 0.49 ± 0.10 |
| *superionic* | 2500 | 175 | 3.39 ± 0.03 | 5.12 ± 0.10 | 21.7 ± 2.5 | 560 ± 50 | 0.73 ± 0.15 |
| *superionic* | 3000 | 175 | 3.36 ± 0.03 | 4.30 ± 0.09 | 14.6 ± 2.2 | 550 ± 50 | 0.59 ± 0.13 |
| *superionic (finite-diff)* | *2750* | 175 | – | 4.66 ± 0.05 | 17.5 ± 0.5 | – | 0.66 ± 0.08 |

**REFERENCES AND NOTES**

1. G. Schubert, D. L. Turcotte, E. R. Oxburgh, Stability of planetary interiors. *Geophysical Journal of the Royal Astronomical Society* **18**, 441-& (1969).
2. D. Gubbins, D. Alfe, G. Masters, G. D. Price, M. J. Gillan, Can the Earth's dynamo run on heat alone? *Geophysical Journal International* **155**, 609-622 (2003).
3. T. Guillot, G. Chabrier, D. Gautier, P. Morel, Effect of radiative transport on the evolution of jupiter and saturn. *Astrophysical Journal* **450**, 463-472 (1995).
4. T. Spohn, G. Schubert, Oceans in the icy Galilean satellites of Jupiter? *Icarus* **161**, 456-467 (2003).
5. W. B. Moore, Thermal equilibrium in Europa's ice shell. *Icarus* **180**, 141-146 (2006).
6. R. Helled, J. D. Anderson, M. Podolak, G. Schubert, Interior models of uranus and neptune. *Astrophysical Journal* **726**, 7 (2011).
7. Y. Kaspi, A. P. Showman, W. B. Hubbard, O. Aharonson, R. Helled, Atmospheric confinement of jet streams on Uranus and Neptune. *Nature* **497**, 344-347 (2013).
8. Z. Alterman, H. Jarosch, C. L. Pekeris, Oscillations of the Earth. *Proceedings of the Royal Society of London Series a-Mathematical and Physical Sciences* **252**, 80-95 (1959).
9. R. Sabadini, R. Vermeersen, *Global Dynamics of the Earth*. (Kluwer Academic, Dordrecht, Netherlands, 2004), pp. 328.
10. I. Matsuyama, M. Beuthe, H. Hay, F. Nimmo, S. Kamata, Ocean tidal heating in icy satellites with solid shells. *Icarus* **312**, 208-230 (2018).
11. H. M. Jara-Orue, B. L. A. Vermeersen, Effects of low-viscous layers and a non-zero obliquity on surface stresses induced by diurnal tides and non-synchronous rotation: The case of Europa. *Icarus* **215**, 417-438 (2011).
12. W. G. Henning, T. Hurford, Tidal heating in multilayered terrestrial exoplanets. *Astrophysical Journal* **789**, 27 (2014).
13. F. Remus, S. Mathis, J. P. Zahn, V. Lainey, The surface signature of the tidal dissipation of the core in a two-layer planet (Research Note). *Astron Astrophys* **573**, 5 (2015).
14. S. Padovan *et al.*, Matrix-propagator approach to compute fluid Love numbers and applicability to extrasolar planets. *Astron Astrophys* **620**, 10 (2018).
15. S. V. Gavrilov, V. N. Zharkov, V. V. Leontev, Influence of tides on the gravitational field of Jupiter. *Soviet Astronomy* **19**, 618-621 (1976).
16. A. S. Nowick, B. S. Berry, *Anelastic relaxation in crystalline solids*. (Academic Press, New York, 1972).
17. J. A. Hernandez, R. Caracas, Superionic-Superionic Phase Transitions in Body-Centered Cubic H2O Ice. *Physical Review Letters* **117**, 5 (2016).
18. J. P. Watt, G. F. Davies, R. J. O. Connell, The elastic properties of composite materials. *Reviews of Geophysics and Space Physics* **14**, 541-563 (1976).
19. H. C. P. Lau *et al.*, A normal mode treatment of semi-diurnal body tides on an aspherical, rotating and anelastic Earth. *Geophysical Journal International* **202**, 1392-1406 (2015).
20. W. G. Henning, R. J. O'Connell, D. D. Sasselov, Tidally heated terrestrial exoplanets: Viscoelastic response models. *Astrophysical Journal* **707**, 1000-1015 (2009).
21. J. C. Castillo-Rogez, M. Efroimsky, V. Lainey, The tidal history of Iapetus: Spin dynamics in the light of a refined dissipation model. *Journal of Geophysical Research-Planets* **116**, 29 (2011).

22. W. C. Tittemore, J. Wisdom, Tidal evolution of the Uranian satellites .2. An explanation of the anomalously high orbital inclination of Miranda. *Icarus* **78**, 63-89 (1989).
23. F. Grasselli, L. Stixrude, S. Baroni, Heat transport in water at icy-giants conditions from ab initio MD simulations. *arXiv*, 2003.12557 (2020).
24. P. Giannozzi *et al.*, Advanced capabilities for materials modelling with QUANTUM ESPRESSO. *Journal of Physics-Condensed Matter* **29**, 30 (2017).
25. P. Giannozzi *et al.*, QUANTUM ESPRESSO: a modular and open-source software project for quantum simulations of materials. *Journal of Physics-Condensed Matter* **21**, 19 (2009).
26. D. Vanderbilt, OPTIMALLY SMOOTH NORM-CONSERVING PSEUDOPOTENTIALS. *Physical Review B* **32**, 8412-8415 (1985).
27. J. M. Sun, B. K. Clark, S. Torquato, R. Car, The phase diagram of high-pressure superionic ice. *Nat. Commun.* **6**, 8 (2015).
28. M. French, T. R. Mattsson, N. Nettelmann, R. Redmer, Equation of state and phase diagram of water at ultrahigh pressures as in planetary interiors. *Physical Review B* **79**, 11 (2009).
29. M. Millot *et al.*, Experimental evidence for superionic water ice using shock compression. *Nat. Phys.* **14**, 297-+ (2018).
30. R. Redmer, T. R. Mattsson, N. Nettelmann, M. French, The phase diagram of water and the magnetic fields of Uranus and Neptune. *Icarus* **211**, 798-803 (2011).
31. N. Nettelmann, R. Helled, J. J. Fortney, R. Redmer, New indication for a dichotomy in the interior structure of Uranus and Neptune from the application of modified shape and rotation data. *Planet Space Sci* **77**, 143-151 (2013).
32. M. Bethkenhagen *et al.*, Planetary Ices and the Linear Mixing Approximation. *Astrophysical Journal* **848**, 9 (2017).